\documentclass[twocolumn, showpacs,preprintnumbers,amssymb,amsmath,aps,prb]{revtex4-1}
\usepackage{graphicx}
\usepackage{bm}
\usepackage{setspace}
\usepackage{natbib}
\usepackage{float}
\usepackage{url}
\usepackage[colorlinks=true, linkcolor=blue, anchorcolor=blue, citecolor=blue, urlcolor=blue]{hyperref}
\begin{document}
\preprint{}

\title{Hard x-ray photoemission and density functional theory study of the internal electric field in SrTiO$_3$/LaAlO$_3$ oxide heterostructures}

\author{E. Slooten$^1$}
\email{e.slooten@uva.nl}
\author{Zhicheng Zhong$^2$}
\author{H.J.A. Molegraaf$^2$}
\author{P.D. Eerkes$^2$}
\author{S. de Jong$^{1,\natural}$}
\author{F. Massee$^{1,\sharp}$}
\author{E. van Heumen$^1$}
\author{M.K. Kruize$^2$}
\author{S. Wenderich$^2$}
\author{J.E. Kleibeuker$^2,\dagger$}
\author{M. Gorgoi$^3$}
\author{H. Hilgenkamp$^2$}
\author{A. Brinkman$^2$}
\author{M. Huijben$^2$}
\author{G. Rijnders$^2$}
\author{D.H.A. Blank$^2$}
\author{G. Koster$^2$}
\author{P.J. Kelly$^2$}
\author{M.S. Golden$^1$}

\affiliation{
$^1$Van der Waals Zeeman Institute, University of Amsterdam, Science Park 904, 1098 XH, Amsterdam, The Netherlands\\
$^2$Faculty of Science and Technology and MESA$^+$ Institute for Nanotechonology, University of Twente, P.O. Box 217, 7500 AE Enschede, The Netherlands\\
$^3$Helmholtz Zentrum Berlin f\"ur Materialien und Energie GmbH, Albert-Einstein-Strasse 15 12489 Berlin, Germany\\
$^\natural$ Current affiliation: SLAC National Accelerator Laboratory, 2575 Sand Hill Road, Menlo Park, CA 94025-7015, United States of America\\
$^\sharp$ Current affiliation: Laboratory of Solid State Physics, Department of Physics, Cornell University, Ithaca, NY 14853, United States of America.\\
$^\dagger$ Current affiliation: Department of Materials Science and Metallurgy, University of Cambridge, Pembroke Street, Cambridge, CB2 3QZ, United Kingdom
}

\date{\today}
\begin{abstract}
A combined experimental and theoretical investigation of the electronic structure of the archetypal oxide heterointerface system LaAlO$_3$ on SrTiO$_3$ is presented. High-resolution, hard x-ray photoemission is used to uncover the occupation of Ti 3d states and the relative energetic alignment - and hence internal electric fields - within the LaAlO$_3$ layer. Firstly, the Ti 2p core level spectra clearly show occupation of Ti 3d states already for two unit cells of LaAlO$_3$. Secondly, the LaAlO$_3$ core levels were seen to shift to lower binding energy as the LaAlO$_3$ overlayer thickness, $n$, was increased - agreeing with the expectations from the canonical electron transfer model for the emergence of conductivity at the interface. However, not only is the energy offset of only $\sim$300meV between $n=2$ (insulating interface) and $n=6$ (metallic interface) an order of magnitude smaller than the simple expectation, but it is also clearly not the sum of a series of unit-cell by unit-cell shifts within the LaAlO$_3$ block. Both of these facts argue against the simple charge-transfer picture involving a cumulative shift of the LaAlO$_3$ valence bands above the SrTiO$_3$ conduction bands, resulting in charge transfer only for $n\ge4$. We discuss effects which could frustrate this elegant and simple charge transfer model, concluding that although it cannot be ruled out, photodoping by the x-ray beam is unlikely to be the cause of the observed behavior. Turning to the theoretical data, our density functional simulations show that the presence of oxygen vacancies at the LaAlO$_3$ surface at the 25\% level reverses the direction of the internal field in the LaAlO$_3$. Therefore, taking the experimental and theoretical results together, a consistent picture emerges for real-life samples in which nature does not wait until $n=4$ and already for $n=2$, mechanisms other than internal-electric-field-driven electron transfer from idealized LaAlO$_3$ to near-interfacial states in the SrTiO$_3$ substrate are active in heading off the incipient polarization catastrophe that drives the physics in these systems.    

\end{abstract}
\maketitle
\section{Introduction}
Since they were first studied in 2004 by Ohtomo and Hwang, SrTiO$_3/$LaAlO$_3$ (STO/LAO) heterostructures have attracted a great deal of interest.\cite{ohtomo2004,Huijben2009,Reyren31082007,doi:10.1146/annurev-conmatphys-062910-140445} The nature of the conduction that occurs when an LAO film four or more unit cells (UC) thick is grown on a TiO$_2$ terminated STO substrate, \cite{S.Thiel09292006} is still a subject of intense debate. The so-called \textquoteleft polar catastrophe\textquoteright \hspace{1pt} is proposed to play a central role in the explanation of this conductivity. \cite{ohtomo2004} According to this model, the electric potential should increase linearly with each layer of LAO that is added to the sample. Because this cannot continue indefinitely, there must be a critical thickness at which the system finds a way of relaxing to a lower energy state. The most widely discussed model entails an electronic reconstruction \cite{Okamoto2004} (electron transfer picture) whereby electrons are transferred from the outermost AlO$_2$ layer to the TiO$_2$ layer at the interface between STO and LAO. Theoretical calculations have predicted a shift of the LAO energy bands of 0.9 eV per monolayer of LAO.\cite{PhysRevLett.104.166804} For LAO overlayers four or more monolayers thick, the top of the LAO valence band should then rise above the bottom of the STO conduction band, leading to the transfer of electrons and averting a potential runaway. The transferred electrons (half an electron per unit cell) then form a conducting quasi two dimensional electron gas (q-2DEG).\cite{Okamoto2004} Despite the simplicity and elegance of this model, as yet such shifts have not been experimentally observed in oxide MBE \cite{PhysRevB.80.241107} or low O$_2$ partial pressure PLD \cite{PhysRevB.84.245124} grown thin films. As the diverging potential, which would certainly exist for an ideal film stack, still needs to be avoided, additional forms of potential compensation or reduction should be considered. One such possibility is the electron doping that would result from oxygen vacancies being formed in the overlayer \cite{PhysRevB.83.205405,PhysRevB.84.245307, PhysRevB.82.165127} or in the substrate.\cite{schneider:192107,PhysRevLett.98.216803,PhysRevLett.98.196802,PhysRevB.75.121404} Removing an oxygen ion from the crystal donates two electrons to the system. These electrons could then occupy the Ti 3d states which are the lowest energy electron-addition states of the whole system, moving to the STO/LAO interface and reducing the built-in potential.
Intermixing of cations around the interface is often mentioned as an alternative to electronic reconstruction,\cite{PhysRevLett.99.155502, 0953-8984-22-31-312201,PhysRevLett.106.036101,Chambers2010} leading to the formation of La$_{1-x}$Sr$_x$Al$_{1-y}$Ti$_y$ \cite{Chambers2010} or Sr$_{1-1.5x}$La$_x$O interfaces.\cite{PhysRevB.85.045401} Note that A-site intermixing by itself does not resolve the polar catastrophe, and it has been pointed out that not diffusive cation disordering but only external (redox) processes can alter the interfacial net charge in systems such as LAO on STO.\cite{0953-8984-23-8-081001} Many of these alternatives to the built-in potential paradigm involve energetics and kinetics that could differ from one film growth laboratory and technique to the next, yet the critical LAO overlayer thickness for metallic behaviour of $n=4$ unit cells seems to be a remarkably robust phenomenon. Understanding this is one of the challenges currently facing the field and makes comparison between results from films grown using different techniques very important.  

In order to fully understand these systems, knowledge of the conductive layer is vital.
The electrostatic fields predicted by the electron transfer picture are so large that they should be easily observable as shifts and (apparent) broadening of LAO core levels on increasing the thickness of the LAO overlayer.\cite{PhysRevB.80.241107} What is measured in a core level experiment is then a weighted total average of core level peaks from each LAO layer. As a result, a potential build-up of 3.2 eV for - for example - 4 layers of LAO would result in significantly broadened La and Al core levels whose maxima would then be shifted by $\sim 1.9$ eV, due to the integration in the experiment over the different layer-by-layer shifts.
Hard x-ray photoemission spectroscopy (HAXPES) is a powerful tool in detecting such core level shifts in buried interfaces.\cite{Fadley2005} Here we present the results of HAXPES experiments on PLD-grown STO/LAO heterointerfaces. We turn our attention in particular to how the LAO core levels behave as the LAO overlayer thickness increases. We present non-zero LAO core level shifts of $\sim 300$ meV between two and six layers of LAO. The experimental data show a built-in potential of the same sign as predicted by simple polarisation models. Its magnitude, however, is one order lower, and not of a unit-cell by unit-cell type. We present a DFT-based model in which 25\% of the oxygen atoms in the top AlO$_2$ layer have been removed. The reduction of potential build-up by oxygen vacancies is shown to be more than sufficient to account for the only modest electrostatic shifts observed in experiment, and would be in keeping with our experimental observation of Ti$^{3+}$ centers already for only $n=2$. We also discuss alternative scenarios, including whether the experiment itself interferes with the build-up or maintenance of the proposed internal field, and although photodoping cannot be ruled out, we present data implying this scenario is unlikely.    

\section{Experiment}
The measurements were performed on STO(001) substrates with LAO overlayer thicknesses $2\leq n\leq6$ UC, where $n$ is the number of LAO layers. In order to investigate the effect of thick LAO overlayers an additional $n=20$ UC sample was grown. The samples are grown by pulsed laser deposition in an oxygen partial pressure of $2 \times 10^{-3}$ mbar. Oxygen partial pressure is known to strongly affect the transport properties of these systems.\cite{Brinkman2007} At this relatively high pressure the presence of oxygen vacancies is believed to be strongly reduced compared to synthesis at lower partial pressures. Details of the sample growth have been published elsewhere.\cite{Huijben2009}

The experiments were performed at the double crystal monochromatized KMC-1 beamline \cite{schaefers:123102} using the HiKE endstation \cite{Gorgoi2009} at Helmholtz Zentrum Berlin. Unless stated otherwise, a photon energy of $h\nu=3$ keV was used, and the total energy resolution was set to be $\sim 500$ meV. \footnote{This enables us to detect core level shifts of the order of $\sim 50$ meV.} No surface treatment was necessary because of the large electron inelastic mean free path $\lambda$ at this photon energy ($\sim$ 4 nm, as calculated by the TPP-2M relation.\cite{Tanuma}) Additional experiments were carried out using a laboratory Al$\colon$K$_{\alpha}$ source (1486 eV, $\lambda=1.9$ nm) and with synchrotron radiation of energies spanning 2.15 keV to 6 keV ($\lambda = 2.8$ nm to $\lambda = 7.4$ nm, respectively).
All experiments were carried out at room temperature and all energy referencing was done by measuring the energy position of Au 4f core levels. Contact between the sample and the gold reference was made using standard silver paint, and - apart from a 2 mm bare space - the film surface was coated with  commercial colloidal graphite contacting agent. During the photoemission experiments themselves - both for the synchrotron-based as well as Al$\colon$K$_{\alpha}$-based - we checked for photon flux dependent effects by varying the photon flux over more than an order of magnitude, confirming that the measured kinetic energy, peak positions and widths remained unchanged. 
In order to eliminate even the smallest error caused by residual electrical charging in the La 4d and Al 2s core level measurements, we determine the difference in binding energy between these core levels and Sr 3d for each sample and use this as our main quantity of interest.  
An inelastic electron background has been subtracted from the data, and they are normalized to have equal area over each of the binding energy regions shown.

\begin{figure}
\includegraphics[width = 8.6 cm]{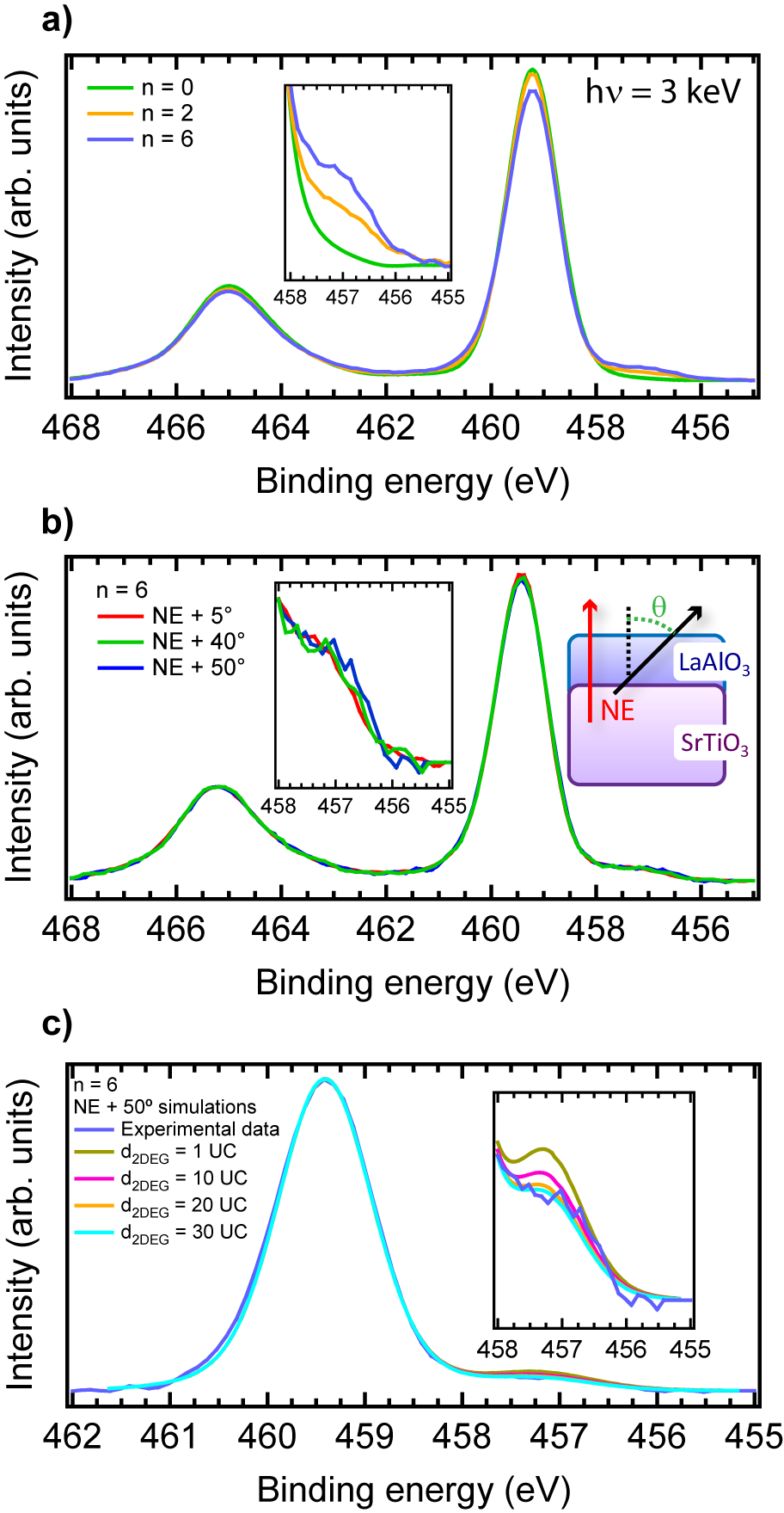}
\caption{(color online) a) The Ti 2p spectrum for $n=2$ and $n=6$. As a reference the spectrum for pure STO ($n=0$) is also included. The shoulder (inset) at 457 eV shows the existence of Ti$^{3+}$ (3d$^1$ entities) for samples both above and below the critical thickness of four layers of LAO. The $n=2$ and the $n=6$ spectra have been shifted to lower binding energy (160 and 220 meV, respectively) in order to align the main line easing comparison between the shoulders. b) The Ti 2p spectrum for the $n=6$ sample as a function of emission angle $\theta$ between 5$^\circ$ and 50$^\circ$ away from normal emission (NE). Within the noise the shoulder (inset) shows marginal or no angular dependence. c) A simulation -  based on the model of Ref. \onlinecite{Sing2009} - of the expected Ti$^{3+}$ shoulder intensity for several degrees of spatial confinement of the q-2DEG for an electron emission angle of 50$^{\circ}$ from normal emission (NE).
\label{fig1}}
\end{figure}

\section{Experimental results and discussion}\label{exp_results}
In Fig. \ref{fig1}a the Ti 2p spectra are shown for a bare STO substrate and the $n=2$ and $6$ samples. As in Ref. \onlinecite{Sing2009} all STO/LAO samples show a shoulder at the low binding energy side of the Ti 2p$_{3/2}$ peak (in our data at 2.3 eV lower binding energy), which is commonly accepted to be a signature of partial occupation of Ti 3d states i.e. the presence of Ti$^{3+}$ ions.
\footnote{The intra-atomic Coulomb interaction between the 2p core hole created as part of the photoemission process and the Ti 3d shell is larger than the band-width of the Ti 3d related states, thus core level photoemission in a system such as this results in two main final states both with a 2p core hole, but one without and one with an electron in the Ti 3d levels.\cite{PhysRevB.54.8446,PhysRevLett.45.1597} The integer 3d electron counts in these two final states are reflected in the constant binding energy difference between the positions of the Ti 2p main lines and shoulder-features, and this underlies the labeling of the shoulders as signaling the presence of `Ti$^{3+}$' ions. Thus, these features in HAXPES do reflect the Ti 3d populations with some fidelity, but do not provide information on the itinerant/localized nature of the electrons in the STO states.}
In order to facilitate comparison of the shoulders in the figure, the $n=2$ and $n=6$ spectra have been shifted to lower binding energy by
160, and 220 meV respectively to coincide with the STO spectrum. The observed absolute Ti 2p$_{3/2}$ binding energy differences for the different LAO thicknesses were similar to those observed in Ref. \onlinecite{PhysRevLett.101.026802}. 
The values of the binding energies of the peak maxima of the Ti 2p$_{3/2}$ and the Sr 3d$_{5/2}$ core levels along with their observed full-width at half-maximum (FWHM) values are given as a function of $n$ in Table \ref{TableI}. We also include the binding energies and FWHM of the relevant LAO core levels. Without going into a blow-by-blow account of all levels for all values of $n$, we do mention that the Ti (Sr) core levels show a significant shift of 160 (240) meV to higher binding energy on addition of LAO on top of the STO substrate. The subsequent increase in the binding energy of these levels as $n$ increases is limited to 70(-20) meV for the Ti 2p (Sr 3d) levels.  
\setlength{\tabcolsep}{7pt}
\begin{center}
\begin{table*}[ht]
\begin{tabular}{ c c c c c}
\hline \hline
\multicolumn{5}{c}{STO core levels}\\
\hline
$n$ & Ti 2p$_{3/2}$ BE  & Ti 2p$_{3/2}$ FWHM  & Sr 3d$_{5/2}$ BE & Sr 3d$_{5/2}$ FWHM \\
  & (eV) & (eV) & (eV) & (eV) \\
\hline
0 & 459.22 & 1.06 & 133.64 & 1.02\\
2 & 459.38 & 1.11 & 133.88 & 1.12\\
3 & 459.43 & 1.13 & 133.90 & 1.11\\
4 & 459.43 & 1.14 & 133.90 & 1.13\\
5 & 459.43 & 1.13 & 133.86 & 1.15\\
6 & 459.45 & 1.14 & 133.86 & 1.16\\
20 & 459.49 & 1.11 & 133.83 & 1.18\\
\hline
\multicolumn{5}{c}{}\\
\multicolumn{5}{c}{LAO core levels}\\
\hline
$n$ & La 4d$_{5/2}$ BE & La 4d$_{5/2}$ FWHM  & Al 2s BE & Al 2s FWHM \\
  & (eV) & (eV) & (eV) & (eV) \\
\hline
2 & 103.20 & 1.84 & 119.55 & 2.00 \\
3 & 103.18 & 1.86 & 119.53 & 1.98 \\
4 & 103.17 & 1.83 & 119.52 & 1.99 \\
5 & 103.04 & 1.89 & 119.39 & 2.08 \\
6 & 102.86 & 1.88 & 119.24 & 2.06 \\
20 & 103.10 & 1.81 & 119.48 & 1.97 \\
\hline \hline
\end{tabular}
\caption{Ti 2p$_{3/2}$, Sr 3d$_{5/2}$ La 4d$_{5/2}$ and Al 2s binding energies (BE) and full width at half maximum (FWHM) as a function of LAO layer thickness. All binding energy values shown have an error bar of $\pm 50$ meV and all widths have an error bar of $\pm70$ meV. All data shown was acquired from spectra measured using h$\nu = 3$ keV and the total energy resolution was 500 meV.}
\label{TableI}
\end{table*} 
\end{center} 
In Fig. \ref{fig1}b, we show that for an $n=6$ film there is no significant dependence of the strength of the Ti$^{3+}$ shoulder on the electron emission angle for angles between 5$^\circ$ and 50$^\circ$ away from normal emission (NE). Similar measurements on samples with other LAO thicknesses yielded the same result.

We simulate the expected Ti$^{3+}$ shoulder using the model adopted in Ref. \onlinecite{Sing2009}: 
\begin{equation}
\frac{I^{3+}}{I^{4+}}=\frac{p(1-e^{-d/(\lambda cos\theta)})}{1-p(1-e^{-d/(\lambda cos\theta)})} \label{Form1}
\end{equation}
Here $\theta$ is the emission angle of the excited photoelectrons, $p$ is the fraction of Ti$^{3+}$ ions in the q-2DEG per unit cell, $d$ is the thickness of the q-2DEG and $\lambda$ is the inelastic mean free path length of 4 nm for the 3 keV photons used. In Fig. \ref{fig1}c we show a simulation of the Ti$^{3+}$ shoulder for an $n=6$ sample for four different thicknesses of the q-2DEG, $\theta = 50^{\circ}$ and $p=0.03$. Comparison of the simulation with the experimental data (see inset) shows that the the confinement of the 3d$^1$ Ti ions (Ti$^{3+}$) is certainly to a region greater than 10 UC, and more likely to 20 UC (8 nm) or more from the interface. This value is a factor of two larger than the upper limit reported in Ref. \onlinecite{Sing2009} for confinement lengths in the $n=6$ samples studied there.
Thus, the main message at this stage is that the Ti$^{3+}$ ions picked up in our core level photoemission experiments are certainly not tightly confined in a 2D layer, for example right at the STO/LAO interface.    

The second clear message our Ti 2p data convey is that the Ti 3d states are partially occupied already at $n=2$ (see Fig. \ref{fig1}a), even though conductivity in transport for the same samples only sets in at $n=4$. The origin of these electrons before the critical LAO thickness for conduction is reached is still unexplained.\cite{Sing2009} Defects, such as oxygen vacancies or adsorbates \cite{PhysRevB.84.245124}
could be the source of these electrons, and the occupation of Ti 3d states for $n<4$ seen here and in Refs. [\onlinecite{PhysRevB.84.245124} and \onlinecite{Sing2009}] point to a transition from localized to delocalized behavior for at least partially pre-existing Ti 3d electrons at $0<n<4$, rather than a transition from zero to finite Ti 3d occupation at the critical LAO thickness.
Seeing as an x-ray beam is an integral part of all core level photoemission experiments, photo-induced electrons are another possibility that need consideration,\cite{Sing2009} and we will return to these issues in Section \ref{photon}. 
Naturally, the presence of Ti$^{3+}$ ions in these systems for $n\geq 4$ is in accordance with the electron transfer model, a model which also predicts large core level shifts in the LAO layers. Thus, in order to investigate the presence of these shifts we now move on to a discussion of the La 4d and Al 2s core levels for all samples, using the Sr 3d level of the STO substrate of the same sample as an internal energy reference.

\begin{figure}
\includegraphics[width = 8.6 cm]{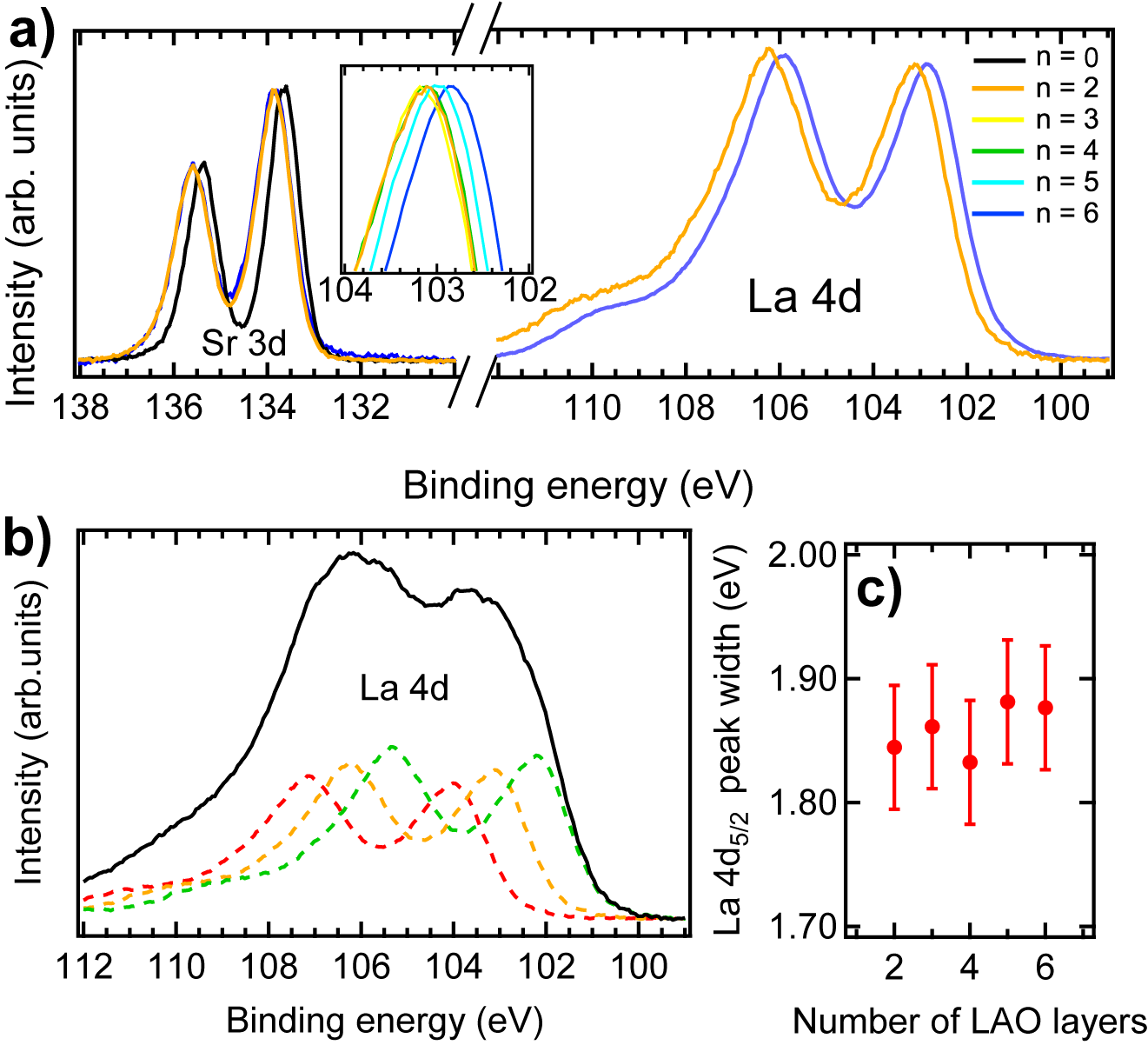}
\caption{(color online) a) La 4d and Sr 3d core level photoemission spectra for selected STO/(LAO)$_n$ samples. The Sr 3d data were measured with $h\nu=2150$ eV, the La 4d data with $h\nu=3000$ eV. The inset shows a clear shift of the La 4d maximum close to 103 eV to lower binding energies with increasing LAO layer thickness, whereas the Sr 3d spectra essentially do not shift in energy between n = 2 and n = 6. b) A simulated $n=3$ spectrum for the La 4d core levels which are shifted by 0.9 eV per LAO unit cell. The colored, dashed spectra are the layer-resolved spectra for the first (interface side, red), second (orange) and third (surface side, green) LAO unit cell, whose $\lambda$-weighted sum then gives the total (black) spectrum. c) The La 4d$_{5/2}$ peak FWHM as a function of LAO layer thickness.} 
\label{fig2}
\end{figure} 

Fig. \ref{fig2}a shows typical La 4d and Sr 3d spectra for $n=2$ and $n=6$. The Sr 3d spectrum for STO ($n=0$) is shown for comparison. As expected, the energy splitting resulting from spin-orbit interaction (Sr 3d$_{5/2}$-3d$_{3/2}$, La 4d$_{5/2}$-4d$_{3/2}$) is unchanged as the LAO overlayer thickness increases. In contrast to the STO core levels for $n\ge2$, however, there is a clear shift for the La 4d peak with increasing LAO thickness. In the inset to Fig. \ref{fig2}a, the La 4d$_{5/2}$ peak maximum is shown for all samples ($n=2-6$), showing the core level shift to lower binding energy in all cases, whereby for $n>4$ this shift is the most striking. The total shift between $n=2$ and $n=6$ amounts to $\sim 300$ meV.

Thus, what Fig. \ref{fig2}a shows is a clear thickness-dependent shift of the LAO core levels to lower binding energy between $n=2$ and $n=6$, that on a qualitative level agrees with the electronic reconstruction model. Having said that, it is also immediately evident that the La 4d core level binding energy changes by much less than the simple expectation of 0.9 eV per LAO unit cell. In Fig. \ref{fig2}b we show the La 4d spectrum simulated for $n=3$, assuming validity of the simple electronic reconstruction model. We use the La 4d HAXPES spectrum from La$_{2-2x}$Sr$_{1+2x}$Mn$_2$O$_7$ with $x=0.425$ as a model lineshape, as we know there is no internal potential build-up in this system\cite{PhysRevB.80.205108}. For $n=3$, the levels in each of the LAO unit cell layers are expected to be shifted by $\sim$ 0.9 eV with respect to the LAO unit cell layer below.\cite{PhysRevLett.104.166804} Consequently, we sum up shifted contributions representing emission from each LAO unit cell layer, and then take an average weighted by $e^{-d/\lambda}$, where $d$ is the depth of the layer and $\lambda$ the electron inelastic mean free path of 4 nm in this case, as this weighted average better represents what could be expected to be measured in experiment.

\begin{figure}
\includegraphics[width = 8.2 cm]{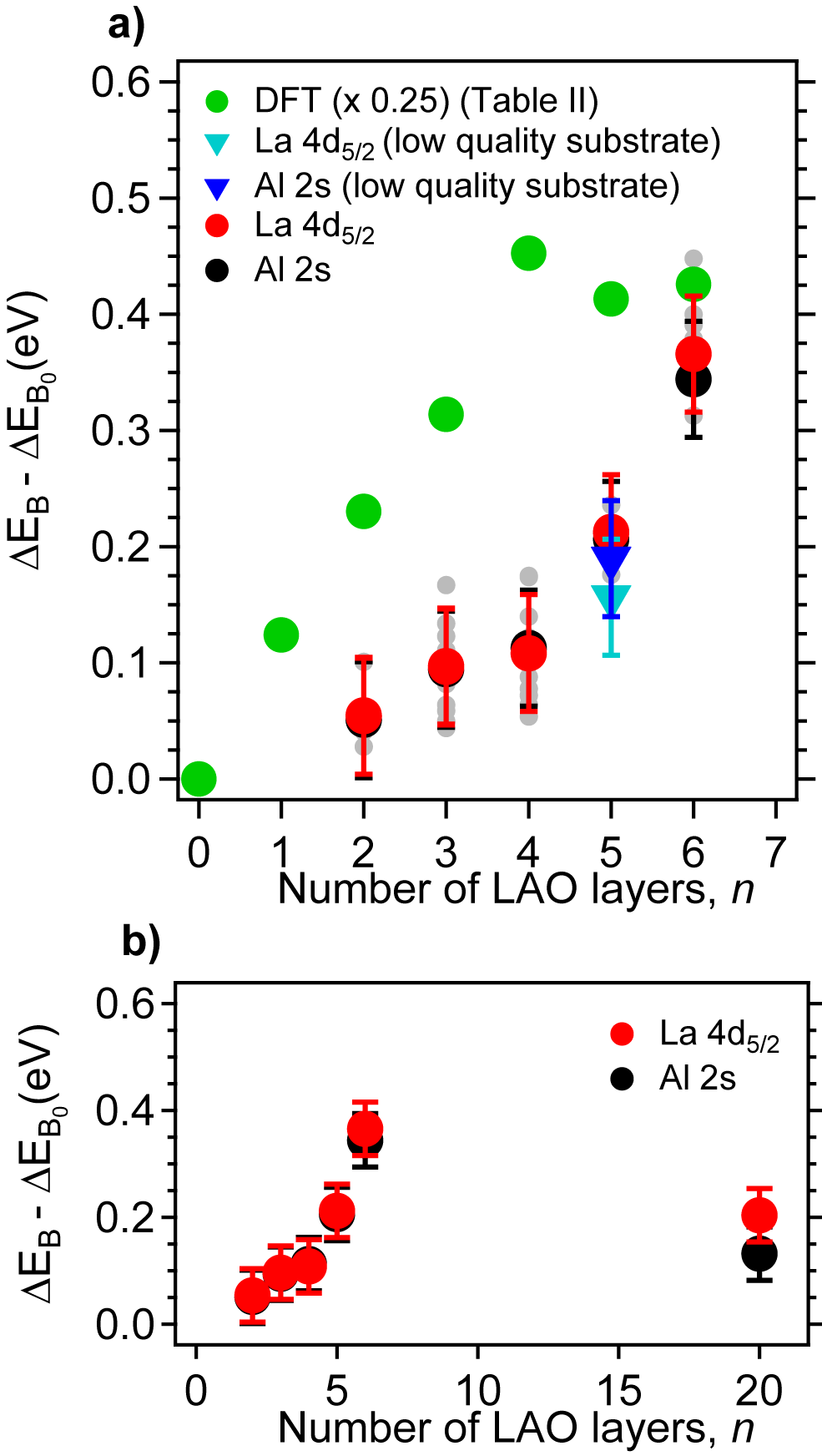}
\caption{(color online) Evolution of the energetics in the LaAlO$_3$ overlayer tracked using the La 4d and Al 2s core level binding energies vs. LaAlO$_3$ thickness, $n$. a) Data up to $n=6$. To indicate the spread in energy shifts encountered, in addition to the main data-points (see legend), the smaller grey symbols show energy shift data from all experiments we conducted on STO/LAO interface samples, including using Al:K$_{\alpha}$ radiation, and using samples from other thin film growth batches.
For comparison with experiment, one fourth of the expected core level shift based on DFT calculations for a vacancy free interface is also shown as an average weighted by the inelastic mean-free path length in HAXPES. The results of these calculations are also shown in Table \ref{TableII}. b) Energy shift data including $n=20$.\label{fig3}}
\end{figure}

It is clear that the simulation in Fig. \ref{fig2}b is too broad - visible for example in the almost complete filling in of the minimum between the two main La 4d peaks at 105 eV $E_B$. The experimentally observed width is essentially unchanged for $n\ge2$, as is shown in Fig. \ref{fig2}c and Table \ref{TableI} \footnote{The expected core level broadening due to the binding energy shifts we observe (Fig. \ref{fig3}) is within the error bars shown in Fig. \ref{fig2}c.}, and there is no indication that the core level peaks consist of a sum of more than one atomic core level spectrum. The data of Fig. \ref{fig2} show that although there is an energy {\it shift} of equal sign to that expected from the electron transfer picture, it differs from the expectations of the simple model in two important respects:
\begin{itemize}
\item we observe a rigid shift rather than a layer-by-layer potential build-up 
\item the shift's magnitude is a (small) fraction of the expected value of $n$ times 0.9 eV/layer from the electronic reconstruction model. 
\end{itemize}

Fig. \ref{fig3} shows the observed core level shifts for all samples discussed here, presented in terms of the binding energy difference between the La or Al core level concerned and the Sr core level from the substrate of the same sample: $\Delta E_B(n)$=$E_{B(n, Sr3d)}$ - $E_{B(n, x)}$ with $x=$ La 4d or Al 2s. In order to be able to plot both the La 4d and Al 2s core level shifts (relative to those of the substrate) on the same scale, the extrapolated value of $\Delta E_B(n)$ to $n=0$, which we denote $\Delta E_{B_0}$ has been subtracted from $\Delta E_B(n)$. The total shift between $n=2$ and $n=6$ amounts to $\sim 300$ meV.
Also shown in Fig. \ref{fig3}a is an $e^{-d/\lambda}$-weighted average of the core level shifts from our DFT calculations for a vacancy free STO/LAO interface, the results of which are shown in Table \ref{TableII}. The details of these calculations are given in Section \ref{DFT}. In order to be able to compare to the results of earlier theoretical predictions,\cite{PhysRevLett.104.166804} we scaled the theoretical core level shifts by a factor of $E_g/E_{g}^{DFT}=3.2/2=1.6$ with $E_g$ being the experimental band gap of STO and $E_{g}^{DFT}$ the DFT value of the STO band gap. Subsequently the data were divided by a factor of four, to fit on the plot shown in Fig. \ref{fig3}a. We note here that a simple `0.9 eV energy shift per unit cell' approximation gives a similar result to the data shown here. 
The comparison between theory and experiment illustrate two things: (i) that the $n$-dependence of the core level shifts observed in experiment is different to the DFT prediction, where in the latter stronger shifts are expected for $n\leq 4$ and essentially no shifts for $n>4$; (ii) the experimental energy-shift data lie well below one fourth of the values predicted by the charge transfer picture. Thus, real-life samples lack both large core level shifts and, as seen in Fig. \ref{fig2}, the connected core level broadening. This forces us to consider that the simple layer-by-layer potential build-up picture is not capturing the main physics at work in real systems. Taken at face value, the severe reduction of the potential build-up manifested by the reduced core-level shifts indicates that other effects, such as defects or reconstructions, should be taken into account.

In Ref. \onlinecite{PhysRevB.80.241107}, XPS data from MBE grown samples recorded using Mg$\colon$K$_{\alpha}$ radiation $h\nu=1253$ eV were reported. There, a total shift $\sim$ 180 meV was observed between samples with $n=2$ and $n=6$ unit cells of LAO. Not only are these shifts 60\% of those we observe, the sign of the shift is opposite to that clearly visible in our raw data. The MBE samples used in Ref. \onlinecite{PhysRevB.80.241107} were grown in an oxygen partial pressure of $10^{-7}$ Torr, so it is certainly conceivable that they contain significantly more oxygen vacancies compared to the \textquoteleft high pressure\textquoteright \hspace{1pt} samples studied here ($p_{O_2}=2\times 10^{-3}$ mbar). We will return to this point later on.

The energy offset we observe between $n=4$ and $n=6$ of $\sim 300$ meV (see Fig. \ref{fig3}) is of the same sign and comparable magnitude to that inferred from recent tunneling transport data on low O$_2$ partial pressure PLD-grown and post-annealed films \cite{Singh-Bhalla2011}. These were observed to enter a Zener tunneling regime at nominally 18.5 unit cells of LAO, which would mean $\sim 300$ meV per unit cell\cite{Singh-Bhalla2011}. Taken at face value, our core level shift data for $n\leq6$ could be argued to be consistent with a picture in which the critical thickness at which the bottom of the STO conduction band is crossed by the top of the LAO valence band would be much greater than $n=6$. A naive extrapolation of our data gives a critical thickness as high as 35 UC. In order to clarify this we measured an $n=20$ sample, whose core level shift data are shown in Fig. \ref{fig3}b. Just as for the $n\leq6$ data, there is no sign of significant broadening of the La and Sr core level lines for $n=20$, ruling out a unit-cell-by-unit-cell energy shift. Furthermore, the binding energy shift for $n=20$ is in fact comparable to that of the $n=5$ sample, illustrating that the LAO energetics probed via the core levels do not simply saturate at some value, but that the behaviour is more complex. In any case, as our experimental data for $n=20$ are inconsistent with the existence of a second critical thickness at high $n$ and with the presence of a step-wise unit-cell by unit-cell potential build-up in the LaAlO$_3$, it would seem unlikely that the picture of an electronic reconstruction at very high $n$ is relevant in these samples.

\section{Photon doping}\label{photon}
At this point it is clear that our data do not display the results of a major unit-cell by unit-cell energy shift. The question arises whether the nature of the experiment itself influences the electrostatics of the sample. The incident hard x-rays used in the experiment create electron-hole pairs that could redistribute in the LAO so as to cancel the potential build-up and therefore flatten the bands. Indeed, it has been observed that any light with an energy greater than the STO band gap dramatically reduces the resistivity of these systems in transport experiments.\cite{Huijben2006, doi:10.1021/nn203991q, PhysRevB.86.075127} 

In the case that x-ray-induced carriers do play a role, they should not only reduce the potential build-up in the LAO overlayer, but should also neutralize the band bending or confining potential for the carriers in the STO substrate. Therefore, if a sufficient number of electron-hole pairs were to be doped into the system in this way, both the potential build-up in the LAO and the band bending in the STO should disappear entirely. In such a scenario - without the confining potential - the Ti 3d electrons present in the LAO/STO system are then free to diffuse through the millimeters thick substrate. Thus, if the system is fully dominated by x-ray-induced electron-hole pair creation, one would expect both no potential build-up in the LAO and no spatial confinement of the Ti$^{3+}$ ions. 

\begin{figure}
\includegraphics[width = 8.6 cm]{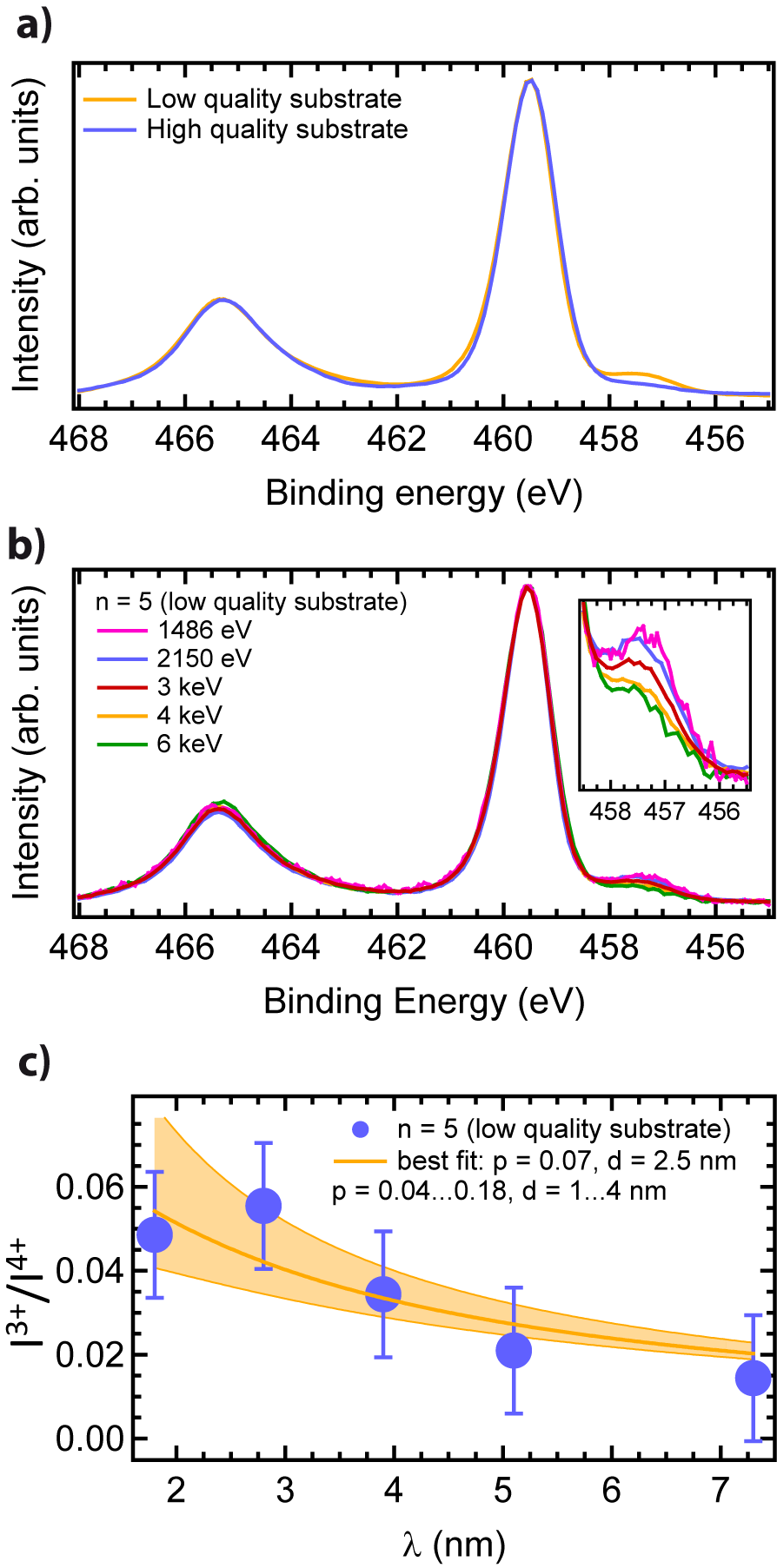}
\caption{(color online) a) The Ti 2p spectrum for the low quality substrate sample compared to the $n=5$ sample from the high quality series. The sample with the lower quality substrate shows a Ti$^{3+}$ shoulder which is clearly much bigger than that of the high quality sample. The spectrum of the low quality substrate was shifted in energy to coincide with that of the high quality film. b) The Ti 2p spectrum for the low quality substrate, $n=5$ sample at different excitation energies between 1486 eV and 6 keV. These spectra have been normalized on the main Ti 2p$_{3/2}$ peak. The intensity of the Ti$^{3+}$ (3$d^1$) shoulder shows a clear kinetic energy dependence, in turn related to the inelastic mean-free path length, indicating confinement of the q-2DEG over 2.5nm, as shown in panel (c) in which the Ti$^{3+}$/Ti$^{4+}$ intensity ratio is plotted as a function of $\lambda$.}\label{fig4}
\end{figure}

In a perfect (unreconstructed), defect-free LaAlO$_3$/SrTiO$_3$ sample where no photon-doping is present at all, the electronic reconstruction scenario would lead to the expected strong potential build-up in the LAO and to confinement of the Ti$^{3+}$ signal to a thin layer of order a few unit cells around the interface. Naturally, between this one extreme and the other, in which photodoping has flattened all potentials and band bending, there is a grey area in which both potential build-up in the LAO and Ti$^{3+}$ confinement both decrease with increasing photon-induced electron-hole pair concentration. One remark is that the band bending confining the Ti 3d electrons is suggested to be of the order of several hundred millivolts,\cite{PhysRevLett.101.026802} whereas the expected potential build-up in the LAO is an order of magnitude greater. Because of this one could expect the Ti$^{3+}$ confinement to be eliminated first as the possible photon-induced electron-hole pair concentration were to be increased. Upon a further increase of hypothetical photo-doping the LAO bands would then flatten further, ultimately resulting in the scenario outlined above.

As discussed above, in the regular samples we see little or no angular dependence of the Ti$^{3+}$ shoulder (see Figs. \ref{fig1}b and c), which - taken at face value - would indicate a thickness of the quasi-2DEG of $\ge8$ nm. However, a second $n=5$ sample was grown under the same conditions as the high quality series, but on a lower quality substrate, which exhibited point defects in an AFM inspection (taken after the usual surface termination procedure, but prior to the LAO deposition). As Fig. \ref{fig4}a shows, the lower quality substrate gives an LAO/STO sample with a large Ti$^{3+}$ contribution: the shoulder to the Ti 2p main line being at least a factor two higher than the one in the regular $n=5$ sample. By varying the kinetic energy of the photoelectrons via changing the photon energy step-wise from 1486 eV to 6 keV a depth profile of the Ti$^{3+}$/Ti$^{4+}$ concentration ratio of the lower quality substrate sample was made. The Ti$^{3+}$/Ti$^{4+}$ intensity ratio extracted from panel (b) of Fig. \ref{fig4} is plotted as a function of $\lambda$ in panel (c). For the special case of normal emission ($\theta = 0$), we can fit Eq. \ref{Form1} to the data and extract a thickness for the q-2DEG of $2.5\pm 1.5$ nm, with the fit parameters for the best fit and the range of parameters that give reasonable fits indicated in the figure legend and as an orange band in Fig. \ref{fig4}c. Both the fit parameters and their error bars are in the same ball park as the ones reported in Ref. \onlinecite{Sing2009}.

If catastrophic photon-induced potential flattening in the LAO were operative under the conditions used for these HAXPES experiments, then the same would presumably hold for the confining potential in the STO, and thus no confinement of the Ti$^{3+}$ to a depth within ca. 2.5 nm of the interface would be observed. Thus, the fact that clear confinement is observed in this particular $n=5$ sample argues against dominance of photon-induced potential flattening in these experiments.
Furthermore, it is germane to note that the La 4d and Al 2s core level binding energies of the lower quality substrate $n=5$ sample are very similar to those of the $n=5$ sample from the regular, high quality series (see Fig. \ref{fig3}a). The fact that these two $n=5$ samples, which were measured in the same photon beam, have such similar La 4d and Al 2s core level binding energies, yet such different Ti$^{3+}$ confinement suggests - but does not prove - that photon doping is not the cause of the lack of band bending in the high quality sample. Which in turn suggests that the level of photon-induced band flattening in the high quality sample series is not overly large.

In an experiment in which we held an $n=3$ sample in the dark for 40 hours and then measured the Sr 3d, Al 2s and La 4d core levels using Al:K$_{\alpha}$ radiation at 10\% of the regular photon flux we observed no change in both the core level binding energies and the core level widths as the measurement progressed over the timescale from 30 minutes to several hours in the beam. Also no change was observed after 24 hours, which would be twice the amount of time required to accumulate 0.5 electrons per UC at the STO/LAO interface at this intensity, as we will discuss in the following, in which we make `back of the envelope' estimations of an upper bound for the number of electron-hole pairs the Al:K$_{\alpha}$ x-ray source used in the lab experiments could generate in the sample. A simple derivation shows this upper bound to be equal to Eq. \ref{Form2}. Assuming that the primary photoelectron current itself corresponds to a negligible amount of the energy deposited in the sample in the form of 1486 eV photons, we reason that each photon represents enough energy to excite across the STO energy gap $1486/3.2=464$ times or 265 times across the LAO energy gap. Given the total photon flux $\phi$ of the Al:K$_{\alpha}$ source of $10^{11}$ photons/s \footnote{Private communication: Dr. Mingqi Wang, VG Scienta}, impinging on a sample area $A$ of $5\times 5$ mm$^2$, and taking the x-ray attenuation length in LAO $\lambda _{LAO}$ to be 0.32 $\mu$m and similarly $\lambda _{STO} = 1.68$ $\mu$m \footnote{Data from: \url{http://henke.lbl.gov/optical_constants/}} we arrive at an area density of electron-hole pairs $n_{EH}$ in an LAO layer of thickness $d_{LAO}=3$ UC of $7.4 \times 10^{10}$ cm$^{-2}$s$^{-1}$.  Naturally, including processes which certainly take place such as energy-loss to phonons would reduce the estimate of the number of possible gap excitations from this upper limit. The number of electron-hole pairs required to cancel the potential build-up is $3.3 \times 10^{14} $cm$^{-2}$ (0.5 electron per UC). Thus, if only electron hole pairs in the LAO contribute to cancellation of the potential build-up, the system would reach this level after some 75 minutes in the Al:K$_{\alpha}$ x-ray beam. Reducing the photon flux by a factor of ten (as we did in the lab experiments) would then increase this time to a little over twelve hours.

\begin{widetext}
\begin{equation}
n_{EH}^{LAO}=\frac{\phi \cdot h\nu}{A \cdot E_{g}^{LAO}}\frac{\lambda _{LAO}(e^{-d_{LAO}/\lambda_{LAO}}-1)}{\lambda _{LAO}(e^{-d_{LAO}/\lambda_{LAO}}-1)+\lambda _{STO}(e^{-d_{LAO}/\lambda _{STO})}} \label{Form2}
\end{equation}
\end{widetext}

Now we consider a scenario in which electron hole pairs created in the top 16 nm (twice the value we found for confinement of the q-2DEG in Fig. \ref{fig1}c) of the STO also contribute to the band flattening in the LAO layer. The additional number of electron-hole pairs created is given by Eq. \ref{Form3}, with $d^*$ the thickness of the STO layer that donates electron-hole pairs. This number amounts to $1.6 \times 10^{12}$ cm$^{-2}$s$^{-1}$, which means one could reach 0.5 electrons per UC already after 3 minutes at full intensity and 30 minutes at 10\% intensity. Given the low photoionisation cross-sections and resulting modest signal to noise ratio in these experiments, it is not possible to determine whether there are any changes in core level binding energies or widths on this time scale. Thus, without dependable information regarding to what extent electron-hole pairs in the STO substrate play a role in possible flattening of the bands, and how much of an overestimate our upper bound represents, these estimations indicate that via this route alone we cannot make a clear call as to the effect of photon doping in these experiments.   
\begin{widetext}
\begin{equation}
n_{EH}^{STO}=\frac{\phi \cdot h\nu}{A \cdot E_{g}^{STO}}\frac{\lambda _{STO}(e^{-d^*/\lambda_{STO}}-e^{-d_{LAO}/\lambda _{STO}})}{\lambda _{STO}(e^{-d^*/\lambda_{STO}}-e^{-d_{LAO}/\lambda _{STO}}) + \lambda _{LAO}(e^{-d_{LAO}/\lambda_{LAO}}-1)+\lambda _{STO}(e^{-d^*/\lambda _{STO})}} \label{Form3}
\end{equation}
\end{widetext}

Before concluding this section dealing with aspects of the experiments themselves that could affect the observed energetics, we briefly discuss sample contacting.\cite{0953-8984-22-31-315501} As related in the experimental section, the surface of the films is additionally contacted by means of colloidal graphite, leaving a strip of 2 mm free on which the x-ray beam impinges. We found this to be an effective manner to prevent charging. A recent paper \cite{PhysRevB.85.125404} uses DFT techniques to study the effect of adding metallic contacts (Ti, Al, Pt) on top of LAO on STO. The conclusions are that such metallic contacts provide a significant charge transfer to the Ti states of the STO, turning them metallic and neutralizing the electric field in the LAO. For Ti and Al these effects are predicted to be largest, as strong covalent bonds between the outermost AlO$_2$ layer of the LAO and the metal result. For Pt, the bonding is weaker and the calculations predict a residual, but reduced electric field in the LAO.\cite{PhysRevB.85.125404}

In our case, a mm away from the region of the films measured using HAXPES there is an overlayer of colloidal graphite. This carbon film can be simply removed by wiping with a tissue impregnated with isopropanol, thus everyday lab practice confirms one's chemical intuition, expecting no covalent bonding and charge transfer between the graphite colloids and the LAO surface. Consequently, one would not expect this adjacent graphite contact layer to affect the energetics observed from the exposed LAO surface from which the data presented here are reported. Indeed: when contacting a sample using a minimal amount of silver conductive paint only on the bottom (substrate side) of the sample we observe no change in any core level binding energies (except due to electrical charging) or peak shape when compared to truly contacting the STO/LAO interface of the sample using the colloidal graphite. 

Before wrapping up as regards the thickness dependent energy shifts observed in LAO on STO from our HAXPES data, it is only fair to mention that a black or white proof either way on the issue of the possible effect of photo-doping or surface photo-voltage effects (and a proof that the sample contacting method plays no role whatsoever) is not possible at this stage, but our considered opinion is that the core-level shifts we report in Figs. \ref{fig2} and \ref{fig3} and their deviation from the simple 0.9 eV unit-cell by unit-cell potential build-up scenario are to be taken seriously.   

\section{Experimental data summary}
Summarizing the experimental results:
\begin{itemize}
\item We have detected a clear low binding energy shoulder as part of the Ti 2p$_{3/2}$ core level photoemission feature, consistent with Ti 3d states being populated with electrons for samples with LaAlO$_3$ thicknesses $n \ge2$. 
\item The relative core level binding energies of the LAO layers decrease as a function of LAO thickness in a manner consistent with an increase in the LAO potential. This apparent potential build-up increases at a higher rate for $n>4$, but decreases again for $n>6$. 
\item Although the sign of the binding energy shifts in LAO agree with the simple electrostatic model for potential build-up, their magnitude - some $\sim$300 meV between $n=2$ and $n=6$  - is much smaller than the predicted 0.9 eV per unit cell.  
\item In addition, the data do not support the existence of a unit-cell by unit-cell alteration of the electrostatic environment in the LAO overlayer, as the LAO core level lines show no measurable increase in width as the overlayer thickness increases.
\end{itemize}

\section{DFT calculations}\label{DFT}
As touched upon earlier, oxygen vacancies could form part of the explanation of both the Ti 3d occupation for $n=2$, and the reduced potential build-up compared to the simple electron-transfer model. Therefore, we turn to an investigation of the effect of oxygen vacancies on core level shifts by means of DFT calculations. In the computer, we have introduced vacancies on 25\% of the oxygen sites in the outermost AlO$_2$ layers of the LAO. As this is the energetically most favorable position for oxygen vacancies in the LAO overlayer \cite{PhysRevB.82.165127} we will only consider this location for oxygen vacancies. For discussion of oxygen vacancies created in other layers of the system as well as in the outermost AlO$_2$ layer, we refer the reader to Ref. \onlinecite{PhysRevB.84.245307}. A 25\% oxygen vacancy concentration would nominally provide enough carriers (half an electron per unit cell) to completely compensate the potential build-up caused by the polar catastrophe, eliminating the need for electronic reconstruction. In this sense, the calculations presented are only a limiting case scenario, but are instructive nonetheless. 

\begin{figure}
\includegraphics[width = 8.6 cm]{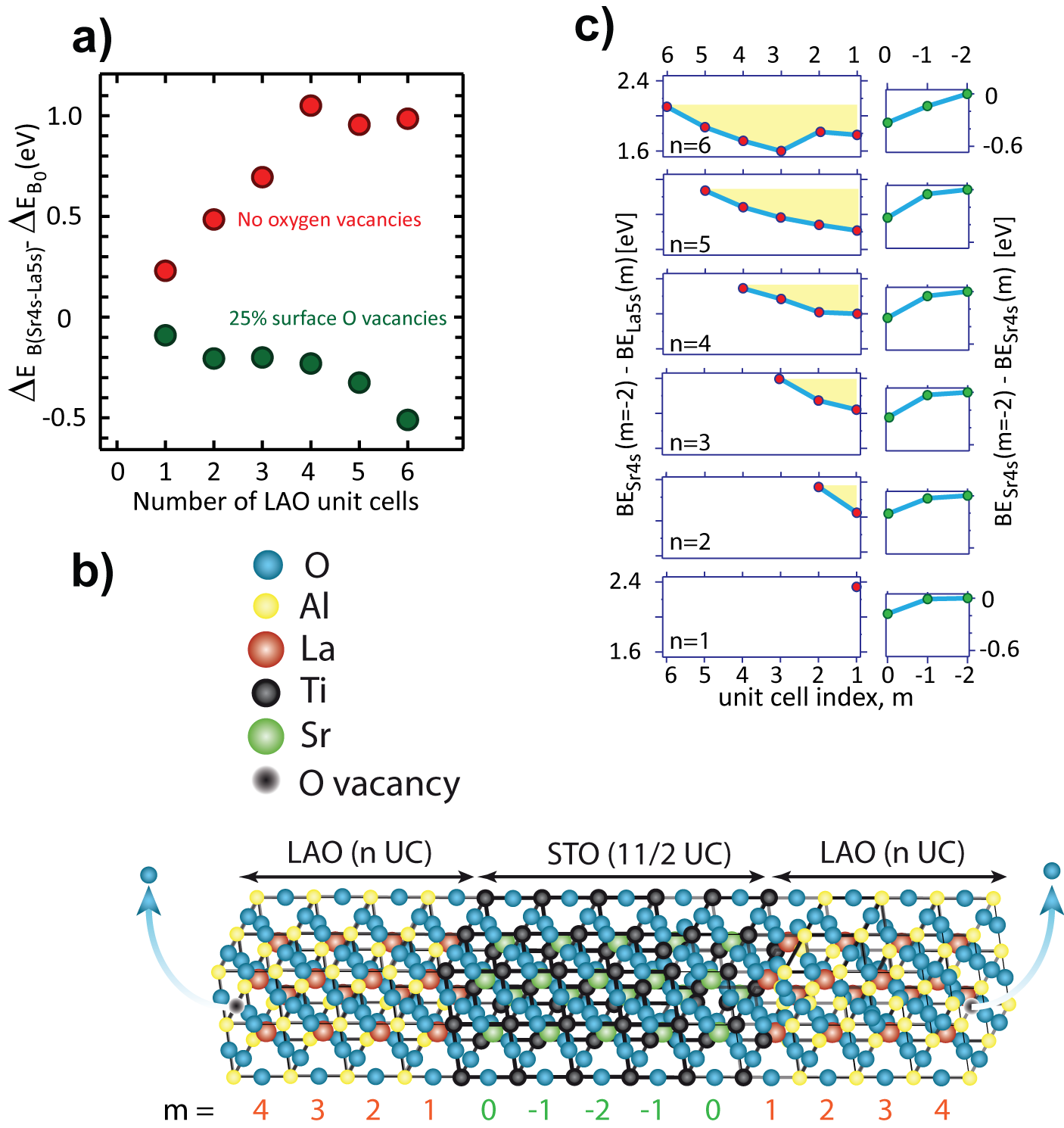}
\caption{(color online). a) The $n$-dependence of the potential in the LaAlO$_3$ expressed via the relative changes in the La 5s shallow core level binding energy extracted from the DFT calculations for vacancy free and oxygen deficient LAO/STO systems described in the text. Points describing a positive slope reflect a trend to decreasing La core level binding energy. b) The $2\times2$ lateral supercell used in the DFT calculations. Oxygen vacancies (25\%) were located in the outermost AlO$_2$ layers. c) Layer-by-layer data from the DFT calculations for the system shown in panel (b). The right hand panels show the $n$-dependence of the energetics in the STO substrate and the left hand panels the evolution of the energies across the LaAlO$_3$ films. The yellow shading indicates regions with LAO potentials such that the La core levels move to higher binding energy, opposite to the direction of the in-built potential scenario.}\label{fig5}
\end{figure}   

For the DFT slab calculations we focus on symmetric ($n$,11/2,$n$) LAO/STO/LAO slabs in which $n$ unit cells of LAO ($2n$ atomic layers) on either side of 11/2 unit cells of STO - resulting in two $n$-type interfaces - are repeated periodically, separated by a vacuum region of approximately 15 \AA, see Fig. \ref{fig5}b. Because the samples are grown on STO substrates, we fix the in-plane lattice constant at the calculated equilibrium value of STO. We introduce oxygen vacancies on both LAO surfaces in such a way that inversion symmetry in the central layer is maintained. Oxygen vacancies at the surface of the LAO are modelled in a 2$\times$2 lateral supercell and all atoms are allowed to relax internally. The initial atomic structure is undistorted, without considering TiO$_6$ and AlO$_6$ octahedron tilting and rotation. Due to symmetry breaking induced by the oxygen vacancies at the surface AlO$_2$ layer, atomic relaxation will lead to an out-of-plane tilting of AlO$_6$ octahedra. \cite{PhysRevB.84.245307} Even so, it still cannot describe a possible in-plane rotation of GdFeO$_3$-like distortions at the interface,\cite{0295-5075-84-2-27001} the investigation of which goes beyond the scope of this article. The computational details have been given elsewhere.\cite{0295-5075-84-2-27001, PhysRevB.82.165127} The effect of strong correlations, which have been shown to be important in STO/RO (R = La, Pr, Nd, Sm) interfaces,\cite{Jang18022011} are not treated explicitly here. Although their inclusion in terms of LDA+U may modify the electrostatic potential profile slightly, it would not change our main theoretical conclusions, which are dominated by the polarity of the system.\cite{PhysRevB.82.165127} The La 4d and Sr 3d core levels convenient for experiment are too deep in binding energy for the calculations, so instead we use the shallower La 5s and Sr 4s core states to track the potential changes at the La and Sr sites. In Figs. \ref{fig5}b and \ref{fig5}c we introduce a unit cell index, $m$, which is $0$ for STO at the LAO interface, negative going deeper into the substrate and $m\ge1$ in the LAO. We calculated the density of states (DOS) for samples with LAO layer thicknesses $1\leq n\leq 6$, projected onto the last few layers of the STO substrate and each of the LAO layers on top. From these data the La 5s and Sr 4s core level binding energies could be extracted, and these are shown in Table \ref{TableII} for the 25\% oxygen vacancies case and for the vacancy free case. The internal energy reference used for both core level energies is the Sr 4s core level in the central SrO layer of the STO substrate furthest from the interface with LAO ($m=-2$).

\begin{table}[ht]
\begin{tabular}{c c c c c c c}
\hline \hline
\multicolumn{7}{c}{25\% oxygen vacancies}\\

\hline
$m$ & 1 UC & 2 UC & 3 UC & 4 UC & 5 UC & 6UC\\
\hline
1 & 2.29 & 2.04 & 2.04 & 2.00 & 1.82 & 1.79\\
2 & - & 2.30 & 2.13 & 2.01 & 1.90 & 1.83\\
3 & - & - & 2.33 & 2.17 & 1.98 & 1.62 \\
4 & - & - & - & 2.34 & 2.09 & 1.74\\
5 & - & - & - & - & 2.33 & 1.90\\
6 & - & - & - & - & - & 2.17\\
\hline
\multicolumn{7}{c}{}\\
\multicolumn{7}{c}{Vacancy free}\\

\hline
$m$ & 1 UC & 2 UC & 3 UC & 4 UC & 5 UC & 6 UC \\
\hline
1 & 2.57 & 2.48 & 2.23 & 2.37 & 2.34 & 2.23\\
2 & - & 3.16 & 2.99 & 2.99 & 2.63 & 2.55\\
3 & - & - & 3.77 & 3.54 & 3.02 & 2.96 \\
4 & - & - & - & 4.36 & 3.54 & 3.30\\
5 & - & - & - & - & 4.45 & 3.78\\
6 & - & - & - & - & - & 4.36\\
\hline \hline
\end{tabular}
\caption{The binding energy difference between Sr 4s ($m=-2$) and La 5s in eV as calculated for both the 25\% oxygen vacancy case and the vacancy free case as a function of LAO thickness in UC for all values of $m$.}
\label{TableII}
\end{table} 

The upper (red) curve in Fig. \ref{fig5}a shows the theory data for the mean-free-path weighted average of the change in the binding energy of the La 5s core level (with respect to an extrapolated value for $n=0$) as a function of the LAO layer thickness (also shown in Fig. \ref{fig3}a, but without the $3.2/2$ bandgap correction). The curve shows a $\lambda$ weighted energy shift equivalent to 0.3 eV per $n$ up to $n=4$ and is then roughly constant thereafter. The per-layer shifts shown in Table \ref{TableII} for the vacancy free case correspond, for instance for the $n=5$ sample, to an average potential build-up of about 0.5 eV per layer and are in agreement with the results of Ref. \onlinecite{Pentcheva2009}. DFT calculations underestimate the magnitude of the STO band gap, putting it at 2.0 eV, whereas the experimental value is 3.2 eV. \cite{Pentcheva2009} This is why - within the electronic reconstruction model - the core levels are in reality expected to shift by 0.9 eV per unit cell, rather than the 0.5 eV per unit cell the DFT data indicate.\cite{PhysRevLett.104.166804} 

Before discussing the lower (green) curve in Fig. \ref{fig5}a, we turn to the layer-resolved core level data from the calculations in Fig. \ref{fig5}c. Such layer-resolved data are inaccessible in a HAXPES experiment, but nonetheless the discussion of this theory data can give some insight into the more detailed processes underlying the type of global, $n$-dependent energy shifts detectable in experiment.

The right hand panels of Fig. \ref{fig5}c indicate that the (near)-interfacial Sr 4s core level binding energy is greater than that in the `bulk' of the STO substrate ($m=-2$) by about 200 meV. This shift is a consequence of the existence of an LaO$^+$ layer immediately atop the TiO$_2$-terminated STO, and the theory data show it to be practically independent of LAO thickness $n$.
This, in fact, agrees with the experimental situation for the Sr 3d and Ti 2p core levels, as described at the outset of Section \ref{exp_results} and shown in Table I. 

The left hand panels of Fig. \ref{fig5}c illustrate the development of the La 5s binding energies (referenced to the Sr 4s of the substrate [$m=-2$]) for the system with 25\% O vacancies. Looking first at the interface ($m=1$), one sees that the binding energy of the La 5s states increases steadily as the LAO thickness increases, leading to a reduction of BE$_{Sr4s}$-BE$_{La5s}$. On the other hand, the STO-referenced La 5s binding energy at the LAO surface (i.e. for $m=n$) remains relatively unchanged as $n$ increases. 

The $\lambda$-weighted sum of the unit-cell by unit-cell changes shown in Fig. \ref{fig5}c gives an overall trend shown in the lower (green) data-points in Fig. \ref{fig5}a. Here, in analogy to the data plotted in Fig. \ref{fig3} we consider the $n$-dependent change in La shallow core level binding energy compared to the hypothetical value for $n=0$, the latter derived from an extrapolation of the $n=6 \rightarrow 1$ values through to $n=0$ (a quantity referred to as $\Delta E_{B_0}$ in the ordinate label). Fig. \ref{fig5}a shows that, for the case of 25\% O vacancies, the La 5s core levels (referenced to a `bulk' Sr 4s level in the center of our slab) shift towards more negative values, indicating a relative increase in the La 5s core level binding energy as $n$ increases for the case of 25\% O vacancies
\footnote {Seeing as in the 25\% oxygen deficient case the DFT data predict the system to be metallic right from the outset, the change from a minor downward shift to a clear movement to higher La 5s binding energy at $n>4$ in Fig. \ref{fig5}a is not linked to the onset of metallicity at $n=4$ observed in transport experiments. The fact that the yellow shaded areas in Fig. \ref{fig5}c shift to the left as $n$ increases, illustrates more La 5s levels with a higher BE than is the case for the calculation for $n=1$, and it is also evident that this proportion increases rapidly for $n>4$. We speculate that one possible cause of this could be an interaction or coupling between holes in the outermost, oxygen deficient surface and electrons at the STO/LAO interface. This interaction could be such that beyond a certain LAO thickness (here $n\sim 4$) the energetics within the LAO nearer the STO substrate changes, resulting in the acceleration in the downward trend seen in the green points of Fig. \ref{fig5}a for $n>4$.}.
This shift is opposite to the vacancy free case (no oxygen vacancies, red symbols in Fig. \ref{fig5}a), and illustrates how the core level shift to lower binding energies expected from the electronic reconstruction scenario can be negated and even turned around by the presence of oxygen vacancies in the LaAlO$_3$.
Because of the non-trivial layer-by-layer changes shown in Fig. \ref{fig5}c, LAO core level binding energies for oxygen vacancy concentrations between zero and 25\% cannot simply be arrived at by appropriate interpolation between the red and green points in Fig. \ref{fig3}a. Nevertheless, the `limiting-case' presented of 25\% O vacancies does illustrate effectively that such defects - and their electronic consequences - will have a strong effect on the energetics in the LAO layers, and are able to neutralize or even reverse the potential build up in the LaAlO$_3$ layer. 

This effect could possibly explain the results of Ref. \onlinecite{PhysRevB.80.241107} as regards the opposite $n$-dependence of the LAO core level binding energies compared to the data presented here, as the samples in Ref. \onlinecite{PhysRevB.80.241107} were grown in a low O$_2$ partial pressure. The experimental data would be consistent with our moderately high O$_2$ partial pressure PLD-grown films being in between the two extremes of the defect-free, pure charge transfer picture and a high oxygen defect scenario. Interpolating between the results from the defect free and 25\% oxygen vacancy calculations may suggest that the observed energy shifts are compatible with an oxygen defect concentration in the top AlO$_2$ layer of our samples of 12-16\%. However, we emphasize that the experimental data show no sign of a unit-cell-by-unit-cell energy offset, whereas - for less than 25\% O defects at the LAO surface - the DFT calculations would predict such a step-wise shift. Because oxygen vacancy formation is driven by the incipient polar catastrophe, this driving force will be diminished upon the formation of vacancies, and the 25\% concentration considered here will be a limiting case. In reality, it is therefore not unreasonable to find a lower oxygen defect concentration and -in addition- other kinds of (charged) defects or non-stoichiometries that also play a role in the flattening out of the built-in potential shifts in real samples.

\section{Conclusion}
In conclusion, we have studied STO/LAO heterointerfaces, grown under moderately high oxygen partial pressure PLD using hard x-ray photoelectron spectroscopy. The aim is the investigation of possible built-in fields in the LaAlO$_3$ by tracking the binding energy differences between the STO and LAO core levels as a function of LAO thicknesses from $n=2$ to $6$ unit cells.

Our data show a positive core level shift with increasing LAO thickness (i.e. binding energies of the LAO core levels decrease): the sign of this shift  meets the expectations from simple built-in potential considerations. The magnitude of the binding energy shift, however, is very low: only $\sim 300$ meV, the majority of which occurs between $n=4$ and $n=6$, and is not of the unit cell by unit cell type. The observed shifts are much lower than the 900 meV per unit cell expected in the simplest ionic polarization models, and remain so even for the large $n$ value of 20.

Our DFT calculations show that oxygen defects in the AlO$_2$ surface layer of the LAO offer one possible explanation for this reduction of the internal potential, as 25\% oxygen defects (sufficient, in principle, to deliver 0.5 electrons per Ti ion at the interfacial layer) are sufficient to even reverse the direction of the energy shift of the La levels in there LAO. 

The facts that we experimentally observe occupation of the Ti 3d states as a shoulder to the Ti 2p signal already for $n=2$ and $n=3$, and that there are only very modest energy shifts observed per additional LAO unit cell strongly suggest that the critical LAO thickness of $n=4$ seen in transport studies is not directly related to a built-in potential-induced shift of the LAO valence band to meet the STO conduction band states. Consequently, it is apparent that in real materials, the electronic reconstruction scenario is pre-empted by other effects, of which we show oxygen vacancies to be one plausible possibility. Influence from photo-induced charge carriers cannot be excluded in our photoionization-based experiments, but we consider it unlikely that these effects dominate the physics we observe. 

Finally, the spectroscopic data do not single out $n=4$ as being special, in contrast to transport data from numerous different groups. This suggests that although the unit-cell by unit-cell potential shift does not - in the end - appear to be the scenario met in real materials, there still must be a mechanism providing `action at a distance', for example from the outermost electrostatic termination layer of the STO/LAO film stack, feeding back to the localized or itinerant character of the states at or near to the STO interface, in this way generating the observed critical thickness seen in transport. 

\section*{Acknowledgments} 
This work is part of the research program of the `Stichting voor Fundamenteel Onderzoek der Materie (FOM)', which is financially supported by the `Nederlandse Organisatie voor Wetenschappelijk Onderzoek (NWO)'. We also acknowledge funding from the EU (via I3 contract RII3-CT-2004-506008 at Helmholtz Zentrum Berlin). This work was supported by ``NanoNed'', a nanotechnology programme of the Dutch Ministry of Economic Affairs and by EC Contract No. IST-033749 "DynaMax". and the Veni program (NWO). The use of supercomputer facilities was sponsored by the ``Stichting Nationale Computer Faciliteiten'' (NCF) which is financially supported by NWO. ES acknowledges S.A. Chambers for useful discussions. 
\bibliography{HAXPES_1st_gen_Bessy_v18}
\end{document}